\documentstyle[aps,eqsecnum]{revtex}
\begin{document}
\draft
\title{Continuous time evolution from iterated maps and Carleman linearization}
\author{P. Gralewicz and K. Kowalski}
\address{Department of Theoretical Physics, University
of \L\'od\'z, ul.\ Pomorska 149/153, 90-236 \L\'od\'z,
Poland}
\maketitle
\begin{abstract}
Using the Carleman linearization technique the continuous iteration
of a mapping is studied.  Based on the detailed analysis of the
Carleman embedding matrix the precise mathematical meaning is given
to such notion.  The ordinary differential equations referring to
continuous iterations are identified and the discussion of the
relationship between them and the corresponding iterated maps is
performed.
\end{abstract}
\section{Introduction}
There are well known examples of how to relate a continuous time
differential system
\begin{displaymath}
\dot x(t)=F(x(t)),
\end{displaymath}
where $F:{\bf R}^k\to{\bf R}^k$, to an iteration of a map of ${\bf
R}^s$ into itself.  We only recall the Euler's broken line method and
the Poincar\'e map.  An example of the inverse procedure to the
discretization has been recently discussed in \cite{1}.  Namely,
a precise meaning has been given therein to the notion of a ``continuous
iteration'' of a mapping, that is a continuous counterpart of the
iterates
\begin{displaymath}
f^n(x)=f(f^{n-1}(x)).
\end{displaymath}

The purpose of this work is to introduce an alternative formalism
for the study of the continuous iterations based on the classical
Carleman linearization technique which is more general than the approach
taken up in \cite{1}.  In section II we briefly introduce the Carleman
linearization technique.  Section III is devoted to the detailed analysis of
the properties of the Carleman embedding matrix which are crucial for the
actual treatment.  Based on the observations of section III we find in
section IV an explicit formula on continuous iterations of a mapping
and show its connection with a linearization transformation for a
corresponding recurrence.  The theory is illustrated by an example of the 
logistic equation.  In section V we introduce the ordinary
differential equation referring to the continuous iteration and find
a simple relation between the Carleman embedding matrices
corresponding to the continuous and discrete time cases.
\newpage
\section{The Carleman linearization}
We begin by recalling the Carleman linearization technique
\cite{2}.  Consider the system
\begin{equation}
\dot x=F(x),
\end{equation}
where $F:{\bf R}^k\to{\bf R}^k$ and $F$ is analytic in $x$.  Having
in mind the applications of the Carleman technique in the study of
the iterated one-dimensional maps discussed in this work we restrict
for brevity to the case with $k=1$, i.e.\ the ordinary differential
equation (2.1).  On making the ansatz
\begin{equation}
x_j := x^j,\qquad j=1,\,2,\,\ldots ,
\end{equation}
where $x$ fulfils (2.1) we arrive at the infinite linear system
\begin{equation}
\dot x_j = \sum_{k=0}^{\infty}L_{jk}x_k,
\end{equation}
with the constant coefficient matrix $L_{jk}$.  Clearly, in view of
(2.2) the finite system (2.1) is embedded into the infinite system
(2.3).  Therefore, the Carleman linearization is also referred to as
the Carleman embedding technique.  Recently, the Carleman approach
has been succesfully applied to the solution of numerous nonlinear
problems (see \cite{3} and references therein).  We only recall the
application of the Carleman linearization technique for calculating
Lyapunov exponents \cite{4} and finding first integrals for the
Lorenz system \cite{5}.

As shown by Steeb \cite{6} the Carleman embedding can be easily
generalized to the case with nonlinear recurrences of the form
\begin{equation}
x_{n+1}=f(x_n),
\end{equation}
where $f$ is analytic in $x_n$.  Indeed, in analogy with (2.2) we set
\begin{equation}
x_{jn} := x_n^j,
\end{equation}
where $x_n$ fulfils (2.4), which leads to the infinite-dimensional
linear system of difference equations such that
\begin{equation}
x_{jn+1}=\sum_{k=0}^{\infty}M_{jk}x_{kn}.
\end{equation}

As with the case of the ordinary differential equations the
finite-dimensional recurrence (2.4) is embedded into the infinite
linear system (2.5).
\section{The Carleman embedding matrix for nonlinear recurrences}
In this section we study the properties of the Carleman embedding matrix
for nonlinear recurrences specified by (2.6) which are utilized in
the actual formalism.  We now return to (2.6).  Let $M(f)$ designate
the Carleman matrix referring to the recurrence (2.4).  Equations
(2.4), (2.5) and (2.6) taken together yield
\begin{equation}
(f(x))^j = \sum_{k=0}^\infty M_{jk}(f) x^k,
\end{equation}
so
\begin{equation}
M_{jk}(f)=\frac{1}{k!}\frac{d^k(f(x))^j}{d x^k}
\bigg\vert_{\raise.5pt\hbox{$\scriptstyle x=0$}}.
\end{equation}
Let
\begin{equation}
f(x) =\sum_{k=0}^\infty f_k x^k.
\end{equation}
The matrix $M(f)$ can be alternatively defined with the help of the
coefficients of the expansion (3.3) as
\begin{equation}
M_{jk}(f) = \left\{
\begin{array}{ll}
\delta_{0k}                                      &\hbox{\hspace{3em}} \hbox{{\rm for }} j=0, \\
\sum\limits_{m_1+m_2+\ldots +m_j=k} f_{m_1} f_{m_2} \cdots f_{m_j} &
\hbox{\hspace{3em}}\hbox{{\rm for }}j\ge1. \\
\end{array}
\right.
\end{equation}
The first few elements of the matrix $M(f)$ are
\begin{equation}
M(f)= \left(
\begin{array}{lllll}
1     & 0           & 0                      & 0
& \ldots \\
f_0   &         f_1 & f_2                    & f_3
& \ldots  \\
f_0^2 & 2 f_0   f_1 & 2 f_0 f_2 + f_1^2      & 2(f_0 f_3+f_1 f_2)
& \ldots  \\
f_0^3 & 3 f_0^2 f_1 & 3(f_0^2 f_2+f_0 f_1^2) &
3f_0^2f_3+6f_0f_1f_2+f_1^3 & \ldots  \\
\vdots  & \vdots         & \vdots                    & \vdots
& \\
\end{array} \right).
\end{equation}
We remark that usage of the formula (3.2) or (3.4) is not the most
effective way of calculating the elements of the matrix $M(f)$.  The
simpler possibility is to apply the relation
\begin{equation}
M_{jk}(f)=\frac{1}{2\pi}\int\limits_{0}^{2\pi}e^{{\rm i}k\varphi}
(f(e^{-{\rm i}\varphi}))^jd\varphi,
\end{equation}
following directly from (3.1) and the well-known fact that the functions
of the form $e^{{\rm i}n\varphi}$, where $n=0,\,1,\,2$,~$\ldots\,$, form the
orthonormal basis of the space of the square integrable functions on
a unit circle.\\[\baselineskip]
\noindent {\em Example:\/} Consider the logistic equation
\begin{equation}
x_{n+1}=\mu x_n(1-x_n).
\end{equation}
Using the relation (3.6) we easily obtain the following formula on
the elements of the corresponding Carleman matrix $M$:
\begin{equation}
M_{jk}=(-1)^{k-j}{j\choose k-j}\mu^j
\end{equation}
(the vanishing of $M_{jk}$ in the case with $2j<k$ is understood).

\vspace{\baselineskip}
We now focus our attention on the mapping
\begin{equation}
f\to M(f).
\end{equation}
A remarkable property of (3.9) is that it provides a representation
of a semigroup of analytic functions with multiplication defined as
the composition operation, that is
\begin{equation}
M(f\circ g)=M(f)M(g).
\end{equation}
The relation (3.10) is an immediate consequence of (3.1).  Clearly,
\begin{equation}
M({\rm id})=I,
\end{equation}
where ${\rm id}(x)\equiv x$ is an identity function playing the role
of a neutral element for the semigroup of the analytic functions,
$I$ is the identity matrix, and whenever exists the inverse $f^{-1}$
of $f$, then
\begin{equation}
M(f^{-1})=M^{-1}(f).
\end{equation}
Finally, it is clear in view of (3.10) that iterations $f^n$ of a
function $f$ are represented by matrix powers, i.e.
\begin{equation}
M(f^n)=M^n(f).
\end{equation}
Having in mind the form of the relation (3.13) it is plausible to
define the continuous iterations $f^t$ of $f$, where $t$ is a real
parameter, by
\begin{equation}
M(f^t)=M^t(f).
\end{equation}
Thus, the problem of the precise definition of continuous iterations
can be reduced to finding the powers $M^t$ of the matrix $M$ for a
non-integer $t$.

We recall that the infinite-dimensional (anti)representations of the
formal power series were originally studied in \cite{1}.  The
counterpart of the formula (3.10) introduced therein describing the
anti-representation is of the form
\begin{equation}
B(f\circ g)=B(g)B(f),
\end{equation}
where $B(f)$ are the Bell matrices specified by
\begin{equation}
B_{jk}(f)=\frac{1}{k!}\frac{d^j(f(x))^k}{d x^k}
\bigg\vert_{\raise.5pt\hbox{$\scriptstyle x=0$}},\qquad
j,\,k=1,\,2,\ldots ,
\end{equation}
and it is assumed that the function $f$ given by the formal power
series satisfies $f(0)=0$.  We point out that such assumption is
rather restrictive one and it is not satisfied in such important
cases as for example $f(z)=z^2+c$, related to the celebrated
Mandelbrot fractal.  Evidently, we have
\begin{equation}
B(f^n)=B^n(f).
\end{equation}
As with (3.13) the relation (3.17) was the point of departure in
\cite{1} to define the continuous iterations.  Nevertheless, the
approach taken up therein is less general and it seems to be more 
complicated than that introduced in the next section of this work.  
We finally remark that in opposition to the actual treatment there 
is no interpretation of the Bell matrices $B(f)$ provided in \cite{1} 
connected with an infinite-dimensional linearization of the original 
nonlinear recurrence (2.4).
\section{Powers of the Carleman embedding matrices and continuous
iterations}
As mentioned in the previous section (see formula (3.14)) the
problem of the precise definition of continuous iterations reduces
to finding the non-integer powers $M^t$ of the matrix $M(f)$.  Our
purpose now is to discuss this point in a more detail.  We first
observe that the problem under investigation can be furthermore cast
into determining the transformation diagonalizing $M$ such that
\begin{equation}
M(f)=U^{-1}\Lambda U,
\end{equation}
where $\Lambda$ is diagonal, i.e.\ $\Lambda_{jk}=\lambda_j\delta_{jk}$.
In fact, (4.1) leads to the following formula on the powers of the
matrix $M$:
\begin{equation}
M^t(f)=U^{-1}\Lambda^t U,
\end{equation}
where $(\Lambda^t)_{jk}=\lambda_j^t\delta_{jk}$.

In order to diagonalize the matrix $M$ we first bring it down to the
triangular form.  Consider (2.4).  Let us assume that $x_n=x_*$ is a
stationary solution to (2.4), that is $x_*$ is a fixed point of $f$
such that
\begin{equation}
f(x_*)=x_*.
\end{equation}
As with ordinary differential equations we can switch over to new
variables
\begin{equation}
x'_n = x_n - x_*,
\end{equation}
so that the resulting nonlinear recurrence
\begin{equation}
x'_{n+1}=g(x'_n),
\end{equation}
where $g(x'_n)=f(x'_n+x_*)-x_*$, obeys
\begin{equation}
g(0)=0.
\end{equation}
Using the definition (3.4) one can easily check that the condition
(4.6) leads to the upper triangular Carleman embedding matrix
$M(g)$.  Further, we have
\begin{equation}
g=h\circ f\circ h^{-1},
\end{equation}
where
\begin{equation}
h(x)=x-x_*.
\end{equation}
Using (3.10), (4.7) and (3.2) we arrive at the matrix relation of
the form
\begin{equation}
M(g)=T_{x_*}M(f)T^{-1}_{x_*},
\end{equation}
where $T_{x_*}=M(h)$, and
\begin{equation}
(T_{x_*})_{ij} = \left\{
\begin{array}{ll}
{j\choose k} (-x_*)^{j-k} &\hbox{\hspace{3em}} \hbox{{\rm for }} j\ge k, \\
0 & \hbox{\hspace{3em}}\hbox{{\rm for }}j<k. \\
\end{array}
\right.
\end{equation}
Evidently,
\begin{equation}
(T^{-1}_{x_*})_{ij} = \left\{
\begin{array}{ll}
{j\choose k} x_*^{j-k} &\hbox{\hspace{3em}} \hbox{{\rm for }} j\ge k, \\
0 & \hbox{\hspace{3em}}\hbox{{\rm for }}j<k. \\
\end{array}
\right.
\end{equation}
We have thus shown that the Carleman embedding matrix $M(f)$
corresponding to (2.4), where $f$ fulfils (4.3) can be
reduced by means of the transformation (4.4) to the upper triangular
form $M(g)$.

Now let
\begin{equation}
g(x) =\sum_{k=0}^\infty g_k x^k.
\end{equation}
Taking into account (3.4) and (4.12) we find that the diagonal
elements of the matrix $M(g)$ are
\begin{equation}
M_{ii}(g)=g_1^i=\left(\frac{dg(0)}{dx}\right)^i=\left(\frac{df(x_*)}{dx}
\right)^i,\qquad i=0,\,1,\,2,\,\ldots .
\end{equation}
These elements coincide with the eigenvalues of the matrix $M(g)$
specified by \cite{7}
\begin{equation}
\psi M(g)=\lambda \psi,
\end{equation}
where $\psi$ is an infinite row-vector.  Suppose now that the eigenvalues
$\lambda_i=M_{ii}(g)$ of the matrix $M(g)$ are mutually different.  Since
$\lambda_i=g_1^i\equiv\lambda^i$, therefore we then have the
restrictive conditions $\lambda \ne0$ and $\lambda\ne\sqrt[n]{1}$.
It is easy to verify that the transformation diagonalizing $M(g)$ of
the form
\begin{equation}
M(g)=V^{-1}\Lambda V,
\end{equation}
where $\Lambda$ is diagonal, is given by the following recursive
relations:
\begin{eqnarray}
V_{jk} = \left\{
\begin{array}{ll}
\displaystyle{(\lambda^j - \lambda^k)^{-1}\sum_{l=j}^{k-1} V_{jl}
M_{lk}(g)} & \hbox{\hspace{3em}}\hbox{{\rm for }}j < k, \\
1 & \hbox{\hspace{3em}}\hbox{{\rm for }}j = k, \\
0 & \hbox{\hspace{3em}}\hbox{{\rm for }}j > k, \\
\end{array}
\right.\\
V^{-1}_{jk} = \left\{
\begin{array}{ll}
\displaystyle{(\lambda^j - \lambda^k)^{-1}\sum_{l=j+1}^k
V^{-1}_{lk}M_{jl}(g)} & \hbox{\hspace{3em}}\hbox{{\rm for }}j < k, \\
1 & \hbox{\hspace{3em}}\hbox{{\rm for }}j = k, \\
0 & \hbox{\hspace{3em}}\hbox{{\rm for }}j > k, \\
\end{array}
\right.
\end{eqnarray}
and
\begin{equation}
\Lambda_{jk}=\lambda^j\delta_{jk}.
\end{equation}
Finally, combining (4.15) and (4.9) we find that the transformation
diagonalizing $M(f)$ can be expressed by (4.1), with
\begin{equation}
U=VT_{x_*}.
\end{equation}

We are now in a position to define the desired continuous iteration
$f^t$.  Indeed, eqs.\ (3.14), (3.1) and (4.1) taken together yield
the following formula on $f^t$:
\begin{eqnarray}
f^t(x) &=& \sum_{k=0}^{\infty}(M^t)_{1k}(f)x^k=\sum_{k=0}^{\infty}
(U^{-1}\Lambda^tU)_{1k}x^k\nonumber\\
&=&\sum_{jklm}(T^{-1}_{x_*})_{1j}V_{jk}^{-1}\lambda^{kt}V_{kl}(T_{x_*})_{lm}
x^m.
\end{eqnarray}
On introducing the functions $\varphi_k(x)$ such that
\begin{equation}
\varphi_k(x) := \left\{
\begin{array}{ll}
\sum_{l=0}^{\infty}V^{-1}_{1k} V_{kl} (x-x_*)^l & \hbox{\hspace{3em}}
\hbox{{\rm for }}k> 0,\\
x_*  & \hbox{\hspace{3em}}\hbox{{\rm for }}k = 0, \\
\end{array}
\right.
\end{equation}
we can write (4.20) in a more compact form.  It follows that
\begin{equation}
f^t(x) = \sum_{k=0}^\infty \lambda^{kt} \varphi_k(x).
\end{equation}

Thus it turns out that the problem of the definition of continuous
iterations can be brought down to the solution of the eigenvalue
equation (4.14).  Further, it is straightforward to show that the
matrix $V$ satisfying (4.15) satisfies the relation (3.4).  Using
this we find that the matrix $V$ can be expressed by the solution to 
the eigenvalue equation (4.14) with the help of the following relation:
\begin{equation}
V_{jk}=\sum_{l=0}^{k}V_{j-1l}\psi_{k-l},
\end{equation}
where $\psi_k$ are the coordinates of the vector $\psi$ .
Now it is not difficult to check that besides of the matrix $V$ also 
the diagonal matrix $\Lambda$ specified by (4.18) fulfils
the relation (3.4).  On passing with the use of (4.1) and (4.19)
from matrices to functions we arrive at the following functional
equation:
\begin{equation}
f(x) = u^{-1}( \lambda u(x)).
\end{equation}
where $u=v\circ h$, and $V=M(v)$.  Clearly, the functional
equivalent of (4.2) is of the form
\begin{equation}
f^t(x) = u^{-1}( \lambda^t u(x)).
\end{equation}
It thus appears that the problem of the definition of the continuous
iterations can be alternatively cast into the solution of the
functional equation
\begin{equation}
u(f(x))=\lambda u(x).
\end{equation}
The equation (4.26) is known in the literature.  For example in
\cite{8} it was used for finding explicit solutions to nonlinear
recurrences (2.4).  We recall that the solution $u(x)$ of (4.26) is
simply the linearization transformation
\begin{equation}
x'_n = u(x_n),
\end{equation}
reducing the solution of the nonlinear recurrence (2.4) to the
linear one
\begin{equation}
x'_{n+1}=\lambda x'_n.
\end{equation}
We point out that the connection of the functional equation (4.26)
with an infinite-dimensional eigenvalue problem was found for the
first time by the second author \cite{9,10} in the context of the Hilbert space
description of nonlinear recurrences (2.4).  The formula on continuous 
iterations analogous to (4.22) in the particular case of $f(0)=0$ was 
originally obtained in \cite{1} with the use of the finite-dimensional 
truncations of the Bell matrices.  We finally remark that our
experience indicates that the recursive relation (4.23) on the matrix
$V$ is of practical importance for the numerical solution of the
functional equation (4.26).\\[\baselineskip]
\noindent {\em Example:\/} Consider as an illustrative example the
well-known exactly solvable case of the logistic equation (3.7) with
$\mu=4$
\begin{equation}
x_{n+1}=4x_n(1-x_n).
\end{equation}
Using the identity
\begin{equation}
\arccos(2x^2-1)=2\arccos x,\qquad 0\le x\le1,
\end{equation}
we get the solution to the functional equation (4.26), where
$f(x)=4x(1-x)$, of the form
\begin{mathletters}
\begin{eqnarray}
\lambda &=& 4, \\
u(x) &=& \hbox{$\scriptstyle 1\over4 $}[\arccos(1-2x)]^2.
\end{eqnarray}
\end{mathletters}
Finally, taking into account (4.25) we get the desired formula on
the continuous iterations such that
\begin{equation}
f^t(x)=\hbox{$\scriptstyle 1\over2 $}\{1-\cos[2^t\arccos(1-2x)]\}.
\end{equation}
Referring back to (4.22) we find
\begin{equation}
x_*=0,\qquad h(x)=x,\qquad \varphi_0=0,\qquad \varphi_k(x)=
\hbox{$\scriptstyle 1\over2
$}(-1)^{k+1}\frac{[\arccos(1-2x)]^{2k}}{(2k!)},\qquad k\ge1.
\end{equation}
Thus the well-known solution to (4.29) of the form
$x_n(x_0)$~=~$f^n(x_0)$, where $f^n(x_0)$ is given by (4.32)
corresponds to the fixed point $x_*=0$ of $f(x)$.  We now discuss
the solution corresponding to the second fixed point
$x_*=\frac{3}{4}$ of $f(x)$.  On utilizing the identity
\begin{equation}
2\pi-\arccos(2x^2-1)=2\arccos x,\qquad -1\le x<0,
\end{equation}
we arrive at the solution to (4.26), where
\begin{mathletters}
\begin{eqnarray}
\lambda &=& -2, \\
u(x) &=& \hbox{$\scriptstyle 1\over2
$}\arccos(1-2x)-\hbox{$\scriptstyle \pi\over3 $}.
\end{eqnarray}
\end{mathletters}
Taking into account (4.25) and (4.35) we obtain
\begin{equation}
f^t(x) = \hbox{$\scriptstyle 1\over2 $}\{1-\cos\{(-2)^t[\arccos(1-2x)
-\hbox{$\scriptstyle 2\pi\over 3$}]+\hbox{$\scriptstyle 2\pi\over
3$}\}\}.
\end{equation}
It can be checked with the use of (4.22), where $x_*=\frac{3}{4}$ and
$h(x)=x-\frac{3}{4}$, that the obtained solution really refers to the
fixed point $x_*=\frac{3}{4}$.  Furthermore, a straightforward
calculation shows that the function (4.32) is equivalent to (4.36)
for non-negative integer $t$, that is the solutions of the logistic
equation (4.29) corresponding to (4.32) and (4.36), respectively,
such that $x_n(x_0)$~=~$f^n(x_0)$, coincide.  Clearly, (4.32) is different 
from (4.36) for non-integer $t$.  We conclude that in the case with the
logistic equation (4.29) the continuous iteration is not unique.
It should be noted that such ambiguity was not recognized in \cite{1}.  
In fact, the solution (4.36) cannot be obtained by means of the approach 
introduced in \cite{1}.  We also remark that uniqueness of the continuous 
iterations referring to (4.29) is violated by the existence of the multiple 
equilibria for (4.29).  It is suggested that it is the case for the general
recurrence (2.4).  We finally point out that due to the term $(-2)^t$ the 
function $f^t(x)$ given by (4.36) is complex-valued.  
\section{From iterated maps to continuous time evolution}
In the previous section we have investigated the continuous
iterations $f^t(x)$ referring to the recurrence (2.4).  Since we can
interpret the continuous parameter $t$ as a ``time variable'',
therefore the question naturally arises on the dynamics of $f^t(x)$.
Consider the continuous iterations $f^t(x)$ and the corresponding
powers of the Carleman embedding matrix $M^t(f)$.  We have
\begin{eqnarray}
(f^t(x))^j &=& \sum_{k=0}^{\infty}(M^t)_{jk}x^k,\\
\frac{d}{dt}M^t &=& \ln M\, M^t.
\end{eqnarray}
Eqs.\ (5.1) and (5.2) taken together yield
\begin{equation}
\frac{d}{dt}f^t(x) = G(f^t(x)),
\end{equation}
where
\begin{equation}
G(x) = \sum_{k=0}^{\infty}(\ln M)_{1k}x^k.
\end{equation}
We have thus shown that the continuous iterations $f^t(x)$ satisfy the
ordinary differential equation (5.3) subject to the initial
condition $f^0(x)=x$, i.e.\ $f^t$ is the flow corresponding to (5.3).
Of course, since $f^t(x)$ is a flow therefore the vector field $G$
can be expressed by $f^t(x)$ with the help of the well-known relation
\begin{equation}
G(x) = \frac{df^t(x)}{dt}\bigg\vert_{\raise.5pt\hbox{$\scriptstyle t=0$}}.
\end{equation}
An alternative form of the vector field $G$ can be obtained directly 
from (4.25).  Indeed, differentiating both sides of (4.25) with respect 
to time we get (5.3) with
\begin{equation}
G(x)=\ln\lambda\, \frac{du^{-1}(u(x))}{dx}u(x).
\end{equation}
Interestingly, there exists a remarkably simple relation between the
Carleman linearization of the differential equation (5.3) and the
Carleman linearization of the original recurrence (2.4).  In fact,
on setting
\begin{equation}
x_j(t) := (f^t(x))^j,
\end{equation}
and using (3.1), (3.14) and (5.2) we arrive at the infinite linear
system
\begin{equation}
\dot x_j = \sum_{k=0}^{\infty}L_{jk}x_k,
\end{equation}
where
\begin{equation}
L = \ln M.
\end{equation}
Thus the logarithm of the Carleman embedding matrix related to the
linearization of the recurrence (2.4) is simply the Carleman
embedding matrix describing the linearization of the differential
equation (5.3).  We finally remark that whenever the solution
$f^n(x)$ to (2.4) is chaotic then it is plausible to expect that the 
solution $f^t(x)$ of the differential equation (5.3) is also chaotic.
As is well-known, the definition of chaotic systems presents a
delicate problem, nevertheless it seems incredible that the
one-dimensional autonomous system (5.3) would exhibit in any sense
the chaotic behavior.  The following example provides a possible
solution of the problem.\\[\baselineskip]
{\em Example:} Consider the continuous iteration (4.32)
corresponding to the solution of the logistic equation (4.29).
An immediate consequence of (5.5) or (5.6) and (4.31) is the following 
differential equation satisfied by $f^t(x)$:
\begin{equation}
\frac{df^t(x)}{dt} = \hbox{$\scriptstyle 1\over2$}\ln2\,\sin
\{\arccos[1-2f^t(x)]\}\,\arccos[1-2f^t(x)].
\end{equation}
Notice that the principal part $\arccos (x)$ of the inverse 
cosine obeys $0\le\arccos(x)\le\pi$, therefore the right-hand side
of (5.10), that is $\frac{df^t(x)}{dt}$, is non-negative.  On the
contrary, in view of (4.32) $f^t(x)$ oscillates.  We conclude that
$f^t(x)$ cannot satisfy (5.10) for arbitrary $t$.  Indeed, an easy
inspection shows that $f^t(x)$ satisfies the following differential
equation:
\begin{equation}
\frac{df^t(x)}{dt} = \widetilde G(t,x,f^t(x)),\qquad f^0(x)=x,\qquad
0\le x\le 1,
\end{equation}
where
\begin{eqnarray}
&&\widetilde G(t,x,f^t(x))\nonumber\\
&&= \left\{
\begin{array}{ll}
\hbox{$\scriptstyle 1\over2$}\ln2\,\sin
\{\arccos[1-2f^t(x)]\}\,\arccos[1-2f^t(x)] &\hbox{\hspace{.5em}} 
\hbox{{\rm for }} 0\le t<\frac{\ln\frac{\pi}{\arccos(1-2x)}}{\ln2},\nonumber\\
-\hbox{$\scriptstyle 1\over2$}\ln2\,\sin
\{\arccos[2f^t(x)-1]\}\{\pi+\arccos[2f^t(x)-1]\} & \hbox{\hspace{.5em}}
\hbox{{\rm for }} \frac{\ln\frac{\pi}{\arccos(1-2x)}}{\ln2}<
t<\frac{\ln\frac{2\pi}{\arccos(1-2x)}}{\ln2},\nonumber\\
\vdots &\hbox{\hspace{.5em}}\vdots\nonumber\\
\hbox{$\scriptstyle 1\over2$}\ln2\,\sin
\{\arccos[1-2f^t(x)]\}\{2k\pi+\arccos[1-2f^t(x)]\} &\hbox{\hspace{.5em}} 
\hbox{{\rm for }} \frac{\ln\frac{2k\pi}{\arccos(1-2x)}}{\ln2}< t <
\frac{\ln\frac{(2k+1)\pi}{\arccos(1-2x)}}{\ln2},\nonumber\\
-\hbox{$\scriptstyle 1\over2$}\ln2\,\sin
\{\arccos[2f^t(x)-1]\}\{(2k+1)\pi+\arccos[2f^t(x)-1]\} & \hbox{\hspace{.5em}}
\hbox{{\rm for }} \frac{\ln\frac{(2k+1)\pi}{\arccos(1-2x)}}{\ln2}<
t<\frac{\ln\frac{2(k+1)\pi}{\arccos(1-2x)}}{\ln2},\nonumber\\
\vdots &\hbox{\hspace{.5em}}\vdots\\
0 & \hbox{\hspace{.5em}}\hbox{{\rm for }}   t=\frac{\ln\frac{l\pi}{\arccos(1-2x)}}{\ln2},\\
0 & \hbox{\hspace{.5em}}\hbox{{\rm for }}   x=0,\hfil(5.12)\\
\end{array}
\right.
\end{eqnarray}
where $k,\,l=1,\,2,\,\ldots$.  Thus, it turns out that (5.10) holds
only for $0\le t<\frac{\ln\frac{\pi}{\arccos(1-2x)}}{\ln2}$.
Nevertheless, in view of (5.12) the form of $\widetilde G$ is
determined completely by (5.10).  In fact, the continuation of
(5.10) expressed by (5.12) is evidently implied by (5.10) and the
fact that the inverse cosine is the infinitely-many-valued function.
It seems that the adequate denomination for (5.10) would be the
``principal part'' of (5.11).  As with (4.29) the differential equation 
(5.11) is also chaotic.  In fact, by (4.32) $0\le f^t(x)\le1$, and the 
Lyapunov exponent $\sigma$ is
\begin{equation}
\sigma =\lim_{t\to\infty}\frac{1}{t}\ln\left|\frac{\partial f^t(x)}
{\partial x}\right|=\ln 2.
\end{equation}
As expected, this exponent coincides with the Lyapunov exponent for
(4.29) \cite{11}.  Notice that the chaoticity of (5.11) does not
contradict the well-known fact based on the Poincar\'e-Bendixon theorem
that the minimal dimension of the phase space of the chaotic
autonomous system is three.  As a matter of fact the nonautonomous
equation (5.11) refers to the two-dimensional phase space with
coordinates $f^t$ and $t$.  Nevertheless, becouse of the non-periodic
dependence of the vector field $\widetilde G$ on the time variable
$t$, the volume of the phase space occupied by the trajectories of
the autonomous system corresponding to (5.11) is, in contrast to
the assumption of the Poincar\'e-Bendixon theorem, infinite one.
\section{Conclusion}
In this work it is shown that the Carleman linearization of
nonlinear recurrences defines the matrix representation of analytic
functions.  Such representation enables a sound definition of
continuous iterations.  On introducing the infinite-dimensional
eigenvalue equation (4.14) related to the problem of the definition
of continuous iterations and translating it back into the language
of the composition of functions, we have arrived at the functional
equivalent (4.26) of (4.14).  As we have mentioned earlier that
equation can be met in the literature.  Nevertheless
its application in the context of continuous iterations as
with (4.25) is most probably new.  In opposition to the alternative
approach introduced by Aldrovandi and Freitas \cite{1} the
interpretation is provided in the actual treatment of the matrices
representing functions in terms of the infinite dimensional
linearization of the original nonlinear recurrence.  We have also
identified the (finite) dynamical system corresponding to continuous
iterations, and we have found a simple formula (5.9) relating the
Carleman embedding matrices in the discrete and continuous time cases.
We note that the naive approach to the continuous iterations
corresponding to (2.4) relying on the formal replacement of the
discrete variable $n\/$ by the continuous one $t\/$ suggests only the
delayed equation $x(t+1)=f(x(t))$ which is equivalent to the
infinite dimensional system of ordinary differential equations.
The simplicity of the approach taken up herein suggests that it
would be a useful tool in the study of nonlinear recurrences as well
as their continuous counterparts.

\end{document}